\documentclass[sigconf]{acmart}
\AtBeginDocument{%
  \providecommand\BibTeX{{%
    \normalfont B\kern-0.5em{\scshape i\kern-0.25em b}\kern-0.8em\TeX}}}

\copyrightyear{2019}
\acmYear{2019}
\setcopyright{rightsretained}

\acmConference{SIGGRAPH '19 Talks}{July 28 - August 01, 2019}{Los Angeles, CA, USA}
\acmDOI{10.1145/3306307.3328180}
\acmISBN{978-1-4503-6317-4/19/07}
\acmBooktitle{SIGGRAPH '19 Talks}

\begin{document}
\settopmatter{printacmref=false} \setcopyright{none}
\renewcommand \footnotetextcopyrightpermission[1]{} \pagestyle{plain}
 
\title{Context-Aware Deep Model for Entity Recommendation System in Search Engine at Alibaba}

\author{Qianghuai Jia}
\email{qianghuai.jqh@alibaba-inc.com}
\affiliation{%
  \institution{Alibaba Group}
  \streetaddress{ }
  \city{Hangzhou}
  \state{}
  \postcode{}
}
\author{Ningyu Zhang}
\email{ningyu.zny@alibaba-inc.com}
\affiliation{%
  \institution{Alibaba Group}
  \streetaddress{ }
  \city{Hangzhou}
  \state{}
  \postcode{}
}
\author{Nengwei Hua} \thanks{Copyright © 2019 for this paper by its authors. Use permitted under Creative Commons License Attribution 4.0 International (CC BY 4.0).}
\authornotemark[1]
\email{nengwei.huanw@alibaba-inc.com}
\affiliation{%
  \institution{Alibaba Group}
  \streetaddress{ }
  \city{Hangzhou}
  \state{}
  \postcode{}
}

\renewcommand{\shortauthors}{Qianghuai Jia, et al.}

\begin{abstract}
 Entity recommendation, providing search users with an improved experience via assisting them in finding related entities for a given query, has become an indispensable feature of today's search engines. Existing studies typically only consider the queries with explicit entities. They usually fail to handle complex queries that without entities, such as "what food is good for cold weather", because their models could not infer the underlying meaning of the input text. In this work, we believe that contexts convey valuable evidence that could facilitate the semantic modeling of queries, and take them into consideration for entity recommendation. In order to better model the semantics of queries and entities, we learn the representation of queries and entities jointly with attentive deep neural networks. We evaluate our approach using large-scale, real-world search logs from a widely used commercial Chinese search engine. Our system has been deployed in ShenMa Search Engine \footnote{m.sm.cn} and you can fetch it in UC Browser of Alibaba. Results from online A/B test suggest that the impression efficiency of click-through rate increased by 5.1\% and page view increased by 5.5\%. 
\end{abstract}




\iffalse
\begin{teaserfigure}
  \includegraphics[width=\textwidth]{sampleteaser}
  \caption{Seattle Mariners at Spring Training, 2010.}
  \Description{Enjoying the baseball game from the third-base
  seats. Ichiro Suzuki preparing to bat.}
  \label{fig:teaser}
\end{teaserfigure}
\fi
\maketitle

\section{Introduction}
\begin{figure}
\centering
\includegraphics [width=0.5\textwidth]{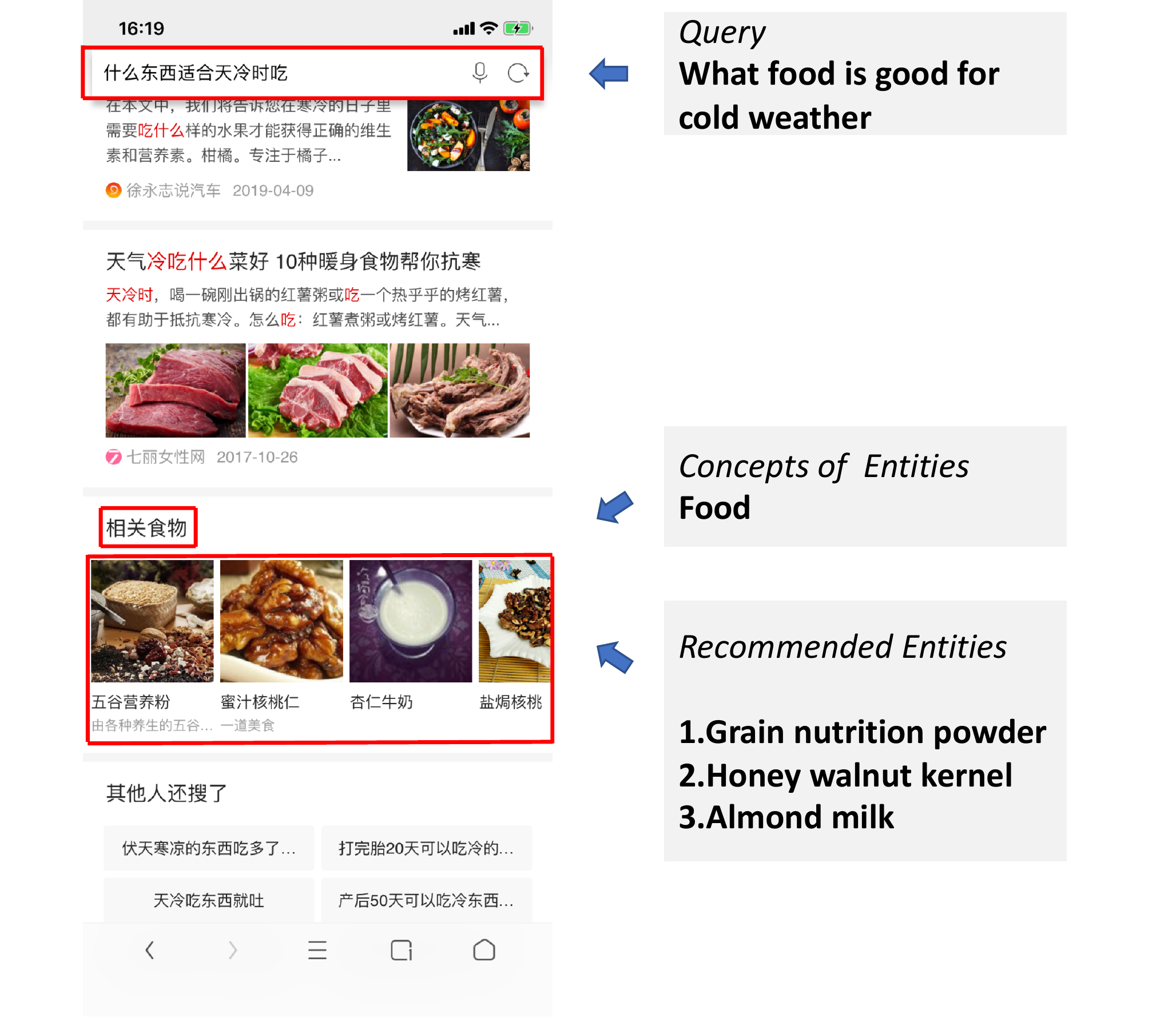}
\caption{Example of  entity recommendation results for the
query "what food is good for cold weather."}
\label{pic1}
\end{figure}

Over the past few years, major commercial search engines have enriched and improved the user experience by proactively presenting related entities for a query along with the regular web search results. Figure \ref{pic1} shows an example of Alibaba ShenMa search engine's entity recommendation results presented on the panel of its mobile search result page.  

Existing studies \cite{blanco2013entity,huang2018improving} in entity recommendation typically consider the query containing explicit entities, while ignoring those queries without entities. A main common drawback of these approaches is that they cannot handle well the complex queries, because they do not have informative evidence other than the entity itself for retrieving related entities with the same surface form. Therefore, existing entity recommendation systems tend to recommend entities with regard to the explicitly asked meaning, ignoring those queries with implicit user needs. Through analyzing hundreds of million unique queries from search logs with named entity recognition technology, we have found that more than 50\% of the queries do not have explicit entities. In our opinion, those queries without explicit entities are valuable for entity recommendation.

The queries convey insights into a user's current information need, which enable us to provide the user with more relevant entity recommendations and improve user experience. For example, a user's search intent behind the query "what food is good for cold weather" could be a kind of food suitable to eat in cold weather. However, most of the entities recommended for the  query are mainly based on entities existed in the query such as given the query "cake" and recommended those  entities "cupcakes," "chocolate" and so on, and there is no explicit entity called "good food for cold weather" at all.  It is very likely that the user is interested in the search engine that is able to recommend  entities with arbitrary queries.   

However, recommending entities with such complex queries is extremely challenging.   At first, many existing recommendation algorithms proven to work well on small problems but fail to operate on a large scale. Highly specialized distributed learning algorithms and efficient serving systems are essential for handling search engine's massive queries and candidate entities. Secondly, user queries are extremely complex and diverse, and it is quite challenging to understand the user's true intention. Furthermore, historical user behavior on the search engine is inherently difficult to predict due to sparsity and a variety of unobservable external factors. We rarely obtain the ground truth of user satisfaction and instead model noisy implicit feedback signals. 

In this paper, we study the problem of context-aware entity recommendation and investigate how to utilize the queries without explicit entities to improve the entity recommendation quality. Our approach is based on neural networks, which maps both queries and candidate entities into vector space via large-scale distributed training. 

We evaluate our approach using large-scale, real-world search logs of a widely used commercial Chinese search engine.  Our system has been deployed in ShenMa Search Engine and you can experience this feature in UC Browser of Alibaba. Results from online A/B test involving a large number of real users suggest that the impression efficiency of click-through rate (CTR) increased by 5.1\%  and page view (PV) increased by 5.5\%. 

The main contributions of our paper are summarized as follows:

\begin{itemize}
\item To the best of our knowledge,  we are the first approach  to recommend entities for arbitrary queries in large-scale Chinese search engine. 
\item Our approach is flexible  capable of recommending  entities for billions of queries.    
\item We conduct extensive experiments on large-scale, real-world search logs which shows the effectiveness of our approach in both offline evaluation and online  A/B test.

\end{itemize}

\section{Related Work}

 Previous work that is closest to our work is the task of entity recommendation. Entity recommendation can be categorized into the following two categories: First,  for query assistance for knowledge graphs \cite{zhang2018attention,zhang2019long}, GQBE \cite{jayaram2014gqbe} and Exemplar Queries \cite{mottin2014exemplar}  studied how to retrieve entities from a knowledge base by specifying example entities. For example, the input entity pair \{Jerry Yang, Yahoo!\} would help retrieve answer pairs such as \{Sergey Brin, Google\}. Both of them projected the example entities onto the RDF knowledge graph to discover result entities as well as the relationships around them. They used an edge-weighted graph as the underlying model and subgraph isomorphism as the basic matching scheme, which in general is costly.
 
 Second, to recommend related entities for search assistance.  \cite{blanco2013entity} proposed a recommendation engine called  Spark to link a user's query word to an entity within a knowledge base and recommend a ranked list of the related entities. To guide user exploration of recommended entities, they also proposed a series of features to characterize the relatedness between the query entity and the related entities. \cite{metzger2013qbees} proposed a similar entity search considering diversity.  \cite{huang2016generating}  proposed to enhance the understandability of entity recommendations by captioning the results.  \cite{fernandez2016memory} proposed a number of memory-based methods that exploit user behaviors in search logs to recommend related entities for a user's full search session.  \cite{huang2018improving} propose a model in a multi-task learning setting where the query representation is shared across entity recommendation and context-aware ranking.  However, none of those approaches take into account queries without entities.   
 
 Our objective is to infer entities given diverse and complex queries for search assistance.  Actually, there are little research papers that focus on this issue. In industry,  there are three simple approaches to handle those complex queries. One is tagging the query and then recommend the relevant entities based on those tags.  However, the tagging space is so huge that it is difficult to cover all domains.   The second method is to use the query recommendation algorithm to convert and disambiguate the queries into entities, ignoring effect of  error transmission from query recommendation.  The last approach is to recall entities  from the clicked documents. However, not all queries have clicked documents. To the best of our knowledge,   we are the first end-to-end method that makes it possible to recommend entities with arbitrary queries in large scale Chinese search engine. 
  \begin{figure}
\centering
\includegraphics [width=0.5\textwidth]{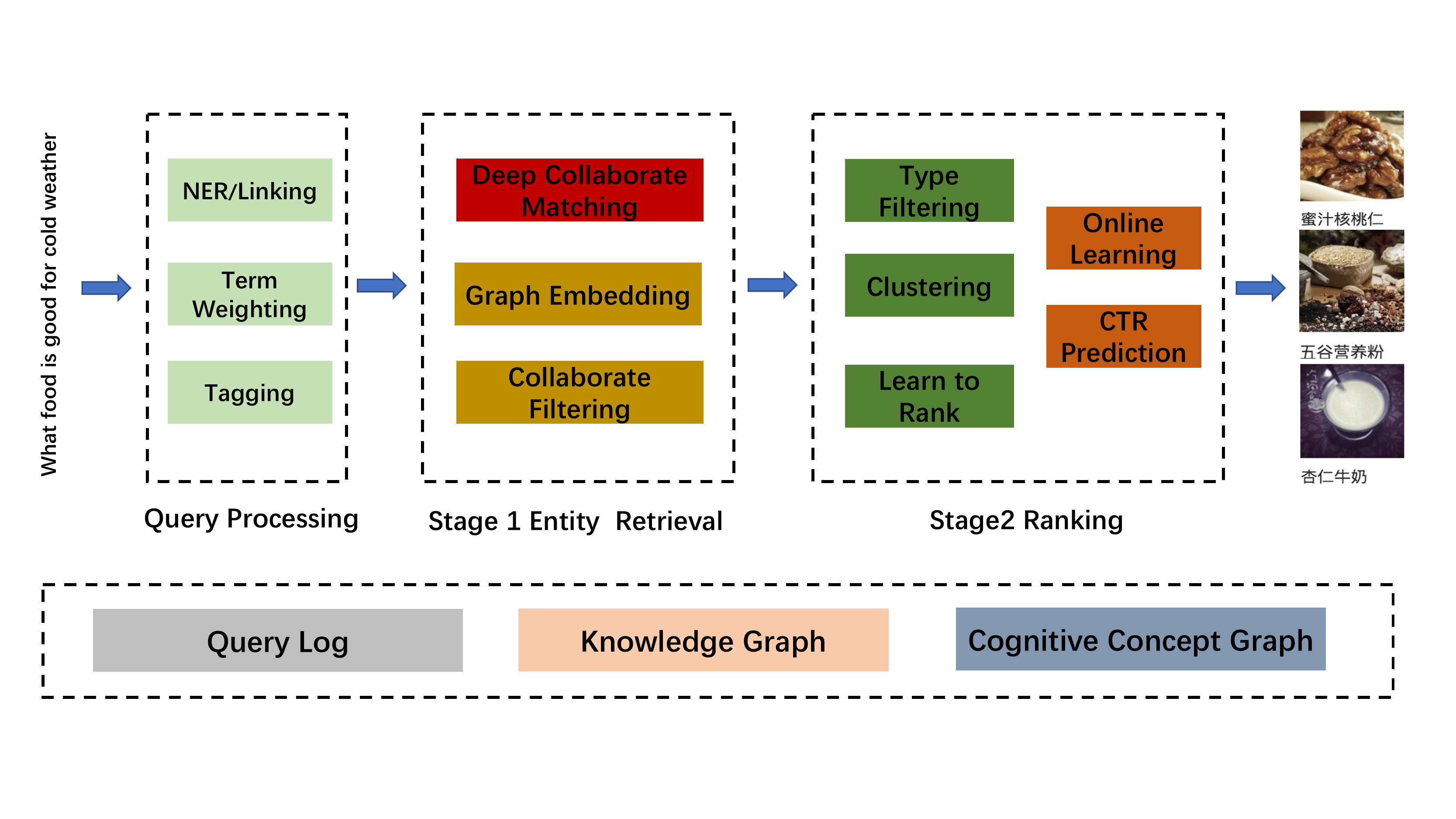}
\caption{System overview  of entity recommendation in ShenMa search engine at Alibaba, the red part is the focus of  this paper.}
\label{arc}
\end{figure}

\section{System Overview}
The overall structure of our entity recommendation system is illustrated in Figure \ref{arc}. The system is composed of three modules: query processing,  candidate generation and  ranking. The query processing module at first  preprocesses the queries, extract entities (cannot extract any entities for complex queries) and  then conceptualize queries.  The candidate generation module takes the output of query processing  module  as input and retrieves a subset (hundreds) of entities from the knowledge graph.  For a simple query with entities, we utilize heterogeneous graph embedding \cite{grover2016node2vec} to retrieve relative entities.   For those complex queries with little entities, we propose a deep collaborative matching model to get relative entities.   These candidates are intended to be generally relevant to the query with high recall. The candidate generation module only provides broad relativity via multi-criteria matching. The similarity between entities is expressed in terms of coarse features.  Presenting a few "best" recommendations in a list requires a fine-level representation to distinguish relative importance among candidates with high precision. The ranking module accomplishes this task by type filtering, learning to rank,  and click-through rate estimation. We also utilize online learning algorithm, including Thompson sampling, to balance the exploitation and exploration in entity ranking.  In the final product representation of entity recommendation,  we utilize the concept of entities to cluster the different  entities with the same concept in the same group to represent a better visual display and provide a better user experience.  In this paper,  we mainly focus on candidate generation, the first stage of entity recommendation  and present our approach (red part in Figure \ref{arc}), which can handle complex queries. 

\section{Preliminaries}
In this section, we describe the large knowledge graph that we use to retrieve candidate entities and cognitive concept graph  that we use to conceptualize queries and entities. 

\subsection{Knowledge Graph}
Shenma knowledge graph\footnote{kg.sm.cn} is a  semantic network that contains ten million of entities, thousand types and billions of triples. It has a wide range of fields, such as people, education, film, tv, music, sports, technology, book, app, food,plant, animal and so on. It is rich enough to cover a large proportion of entities about worldly facts. Entities in the knowledge graph are connected by a variety of relationships.  
\subsection{Cognitive Concept Graph}
Based on Shenma knowledge graph, we also construct a cognitive concept graph which contains millions of instances and concepts. Different from Shenma knowledge graph,    cognitive concept graph is a probabilistic graph mainly focus on the Is-A relationship.   For example, "robin" is-a bird, and "penguin" is-a bird. Cognitive concept graph is helpful in entity conceptualization and query understanding. 

\section{Deep Collaborative Match}
In this section, we first introduce the basics of the deep collaborative match and then elaborate on how we design the deep model architecture. 
\subsection{Recommendation as Classification}
Traditionally, major search engines recommend related entities based on their similarities to the main entity that the user searched. \cite{huang2018improving} have detailed explained the procedure of entity recommendation in the search engine, including entity linking, related entity discovery and so on. Unlike traditional methods, we regard recommendation as large-scale multi-classification where the prediction problem becomes how to  accurately classify a specific entity $e_{i}$ among millions of entities from a knowledge graph  $V$ based on a user's input query $Q$,
$$P(e_{i}|Q)=\frac{u_{i}q}{\sum_{j\in V}u_{j}q}$$
where $q\in \mathbb{R}^{N}$ is a high-dimensional "embedding" of the user's input query, $u_{j}\in \mathbb{R}^{N}$ represents each entity embedding and V is the entities from knowledge graph. In this setting, we map the sparse entity or query into a dense vector in $\mathbb{R}^{N}$. Our deep neural model tries to  learn the query embedding via the user's history behavior  which  is useful for discriminating among entities with a softmax classier. Through joint learning of  entity embeddings and query embeddings, the entity recommendation becomes  the calculation of cosine similarity between entity vectors and query vectors.  
\subsection{Base Deep Match Model}
\begin{figure}
\centering
\includegraphics [width=0.5\textwidth]{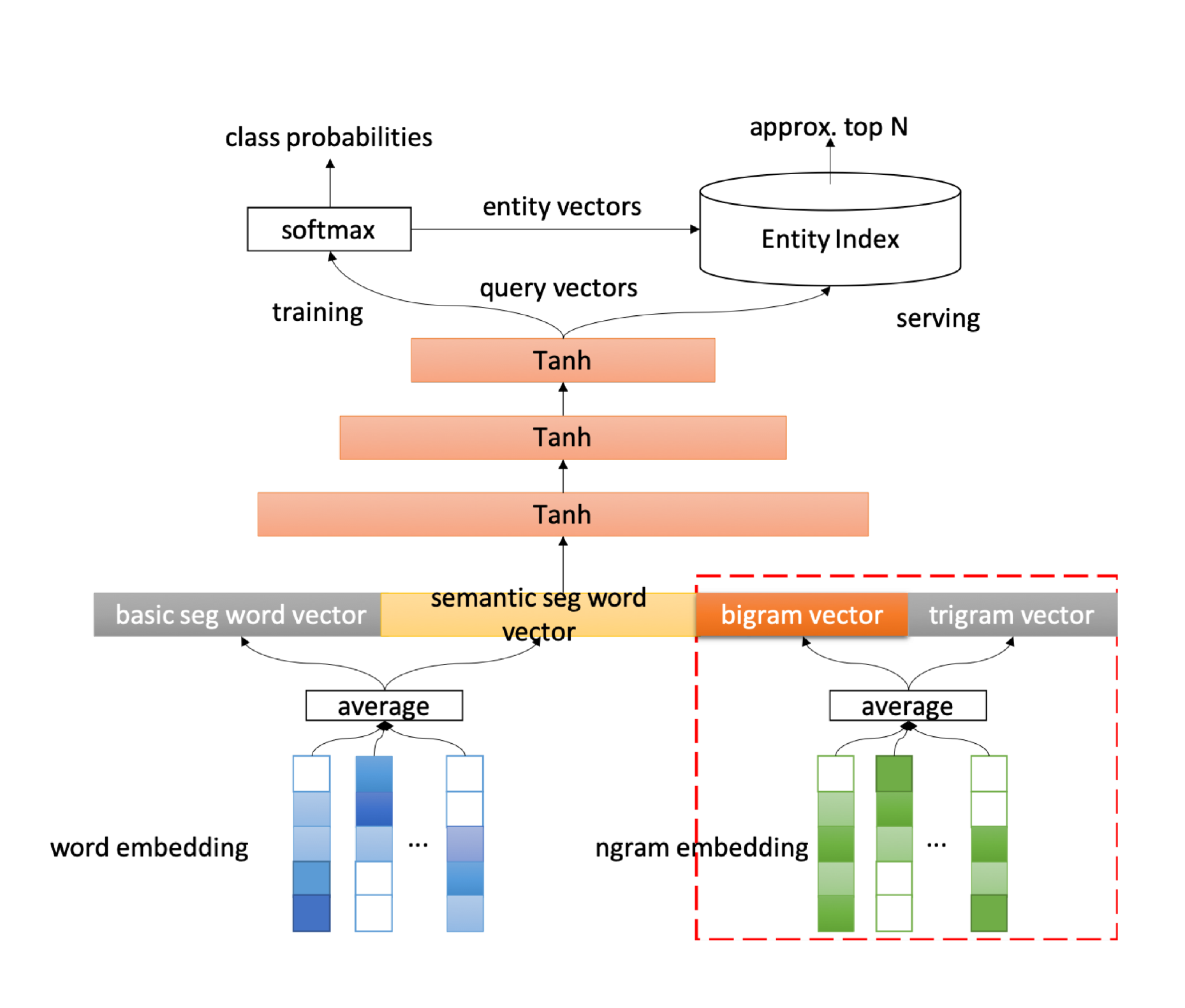}
\caption{Base deep match model.}
\label{base}
\end{figure}
Inspired by skip-gram language models \cite{mikolov2013distributed}, we map the user's input query to a dense vector representation and learn high dimensional embedding for each entity in a knowledge graph. Figure \ref{base} shows the  architecture of the base deep match model.

\textbf{Input Layer.}
Input layer mainly contains the features from the input query, we first use word segmentation tool\footnote{AliWS, which is similar to jieba segmentation tool and uses CRF and user-defined dictionary to segment queries.} to segment  queries,  then fetch basic level tokens and semantic level tokens\footnote{Tokens that in the same entity or phrase will not be segmented.}, and finally combine all the input features via the embedding technique, as shown below:
\begin{itemize}
\item \textbf{word embedding}: averaging the  embedding of both the  basic level tokens and semantic level tokens, and the final embedding dimension is  128. 
\item \textbf{ngram embedding}: inspired by fasttext \cite{joulin2016fasttext}, we add ngram (n=2,3) features to the input layer to  import  some local temporal information. The  dimension of ngram embedding is  also  128.
\end{itemize}

\textbf{Fully-Connected Layer.}
Following the input layer,  we  utilize three fully connected layers (512-256-128) with tanh activation function. In order to speed up the training, we add batch normalization to each layer.

\textbf{Softmax Layer.}
To efficiently train such a model with millions of classes, we apply sampled softmax \cite{blanc2017adaptive} in our model. For each example, the cross-entropy loss is minimized for the true label and the sampled negative classes. In practice,  we sample 5000 negatives instances.

\textbf{Online Serving.}
At the serving time, we need to compute the most likely $K$ classes (entities) in order to choose the top $K$ to present to the user. In order to recall the given number of entities within ten milliseconds, we deploy the vector search engine\footnote{The vector search engine is similar to the facebook's faiss vector search engine, and optimized  in the search algorithm.} under the offline building index. In practice, our model can generate query embedding within 5ms and recall related entities within 3ms.
\subsection{Enhanced Deep Match Model}

\begin{figure}
\centering
\includegraphics [width=0.5\textwidth]{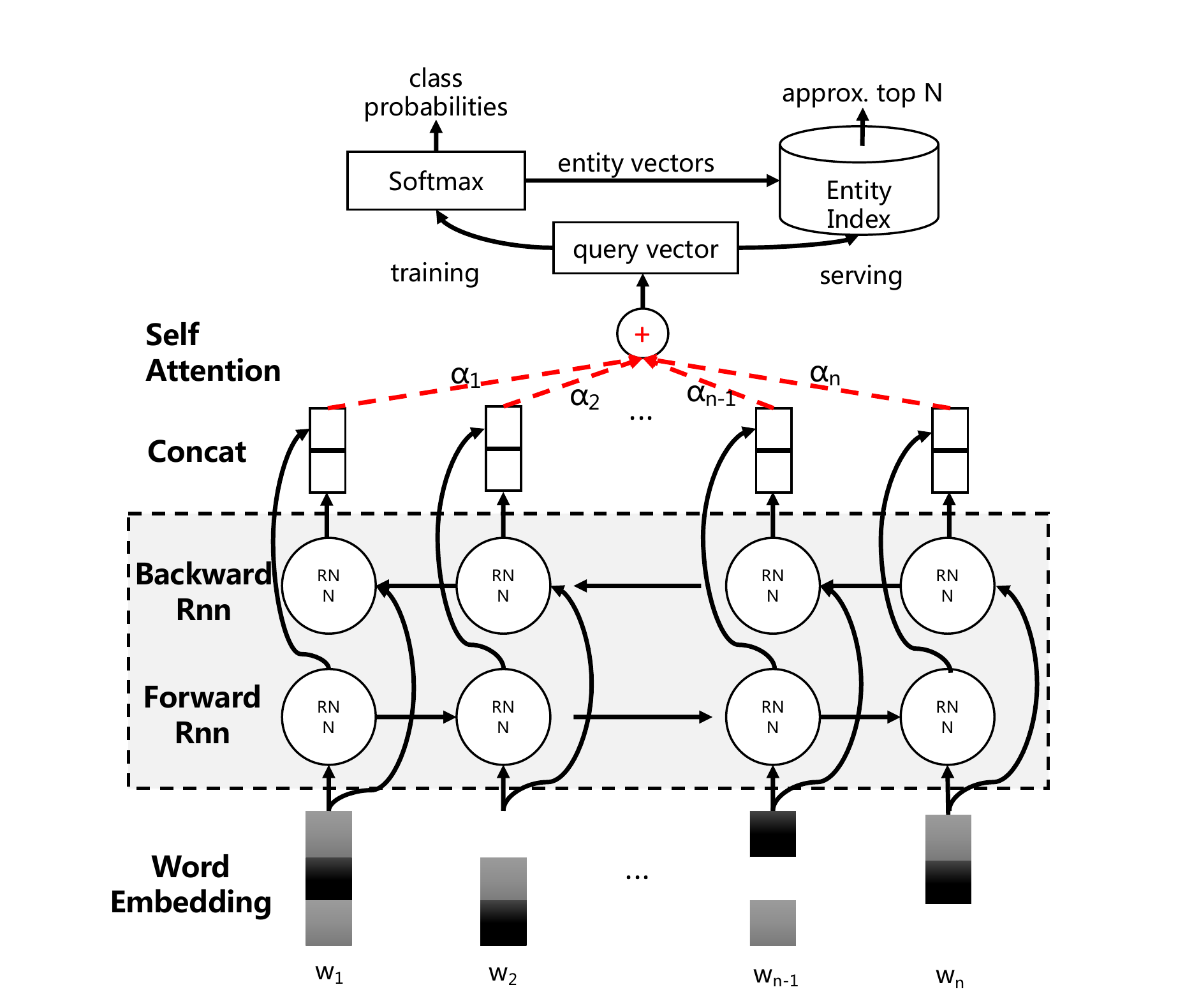}
\caption{Enhanced deep match model.}
\label{pic2}
\end{figure}

The above base model also remains two problems of on the semantic representation of the input query: 1) ignoring the global temporal information, which is important for learning query's sentence-level representation;  2) different query tokens contribute equally to the final input embedding, which is not a good hypnosis. For example, the entity token should be more important than other  tokens such as stop words. 

To address the first issue, we adopt the Bi-directional LSTM model to encode the global and local temporal information. At the same time, with the attention mechanism, our model can automatically learn the weights of  different query tokens. Figure \ref{pic2} shows the enhanced deep match model architecture.

The proposed model consists of two parts. The first is a Bi-directional LSTM, and the second is the self-attention mechanism, which provides weight vectors for the LSTM hidden states. The weight vectors are dotted with the LSTM hidden states, and the  weighted LSTM hidden states are considered as an embedding for the input query. Suppose the input query has $n$ tokens  represented  with  a sequence of word embeddings.
$$Q=(w_{1},w_{2},\cdots ,w_{n-1},w_{n})$$
where  $w_{i}\in \mathbb{R}^{d}$ is the word embedding for the $i$-th token in the query. $Q\in \mathbb{R}^{n \times d}$ is thus represented as a 2-D matrix, which concatenates all the word embeddings together. To utilize the dependency between adjacent words within a single sentence, we use the  Bi-directional LSTM to represent the sentence and concatenate $h_{if}$ with $h_{ib}$ to  obtain the  hidden state $h_{i}$:
$$h_{i}=[h_{if},h_{ib}]$$
The number of    LSTM's  hidden unit  is  $m$. For simplicity, we concatenate all the hidden state $h_{i}$ as $H \in \mathbb{R}^{n \times 2m}$.
$H=[h_{1},h_{2},\cdots ,h_{n-1},h_{n}]$
With the self-attention mechanism,  we encode a variable length sentence into a fixed size embedding. The attention mechanism takes the whole LSTM hidden states $H$ as input, and outputs the  weights $\alpha \in \mathbb{R}^{1 \times k}$:
$$\alpha=softmax(Utanh(WH^{T}+b))$$
where  $W\in \mathbb{R}^{k \times 2m}$,$U\in \mathbb{R}^{1 \times k}$,$b\in \mathbb{R}^{k}$.
Then we sum up the LSTM hidden states $H$ according to the weight provided by $\alpha$ to get the  final
representation  of the input query.
$$q=\sum_{i=1}^{n}\alpha_{i}  h_{i}$$

Note that,  the query embeddings and entity embeddings are all random initialized  and  trained from scratch. We have huge amounts of training data which is capable of modeling the relativity between queries and entities.

\section{Experiments}
\subsection{Data Sets}
In this section, we illustrate how to generate the training samples to learn the query-entity match model. Training samples are generated from query logs and knowledge graph, which can be divided into four parts as  shown below:
\begin{itemize}
\item \textbf{Query-Click-Entity}: given a query, choose the clicked entities with relatively high CTR. In practice, we collect thousand millions of data from the query logs in the past two months.
\item \textbf{Query-Doc-Entity}: we assume that high clicked doc is well matched to the query and the entities in title or summary are also related to the query. The procedure is 1) we first fetch the clicked documents with title and summary from the query log; 2) extract entities from title and summary via name entity recognition; 3) keep those high-quality entities. At last, we collect millions of unique queries.
\item \textbf{Query-Query-Entity}: given the text recommendation's well results, we utilize the entity linking method to extract entities from those results. We also collect millions of unique queries.
\item \textbf{Query-Tag-Entity}: as to some specific queries, we will tag entity label to them and  generate query-entity pairs. Here, we define hundreds of entity tags in advance.
\end{itemize}
After generating of query-entity pairs, we adopt the following data prepossessing procedures:

\begin{figure*}
\centering
\includegraphics [width=0.8\textwidth]{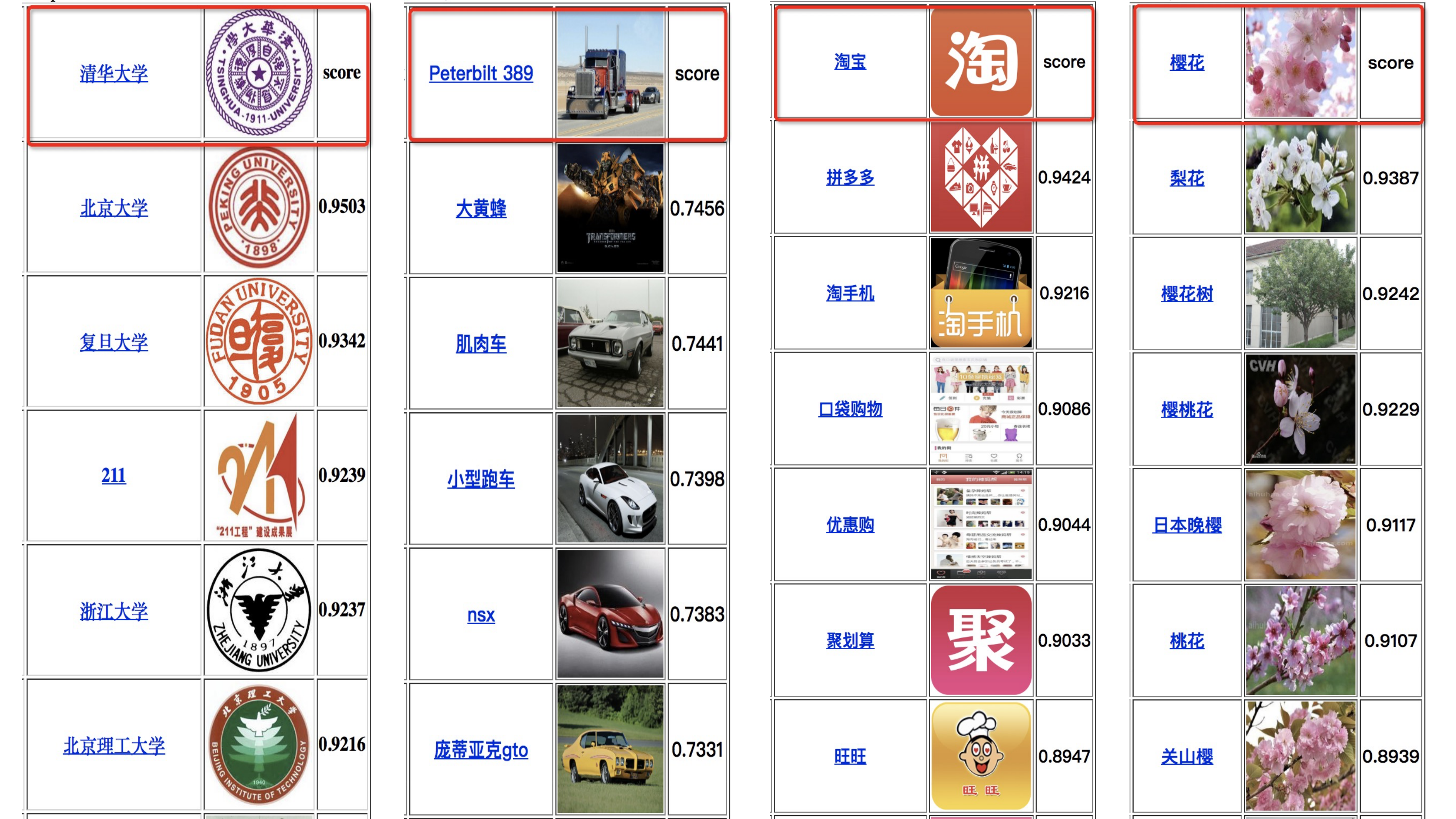}
\caption{The top-N similar entities  for  given entities via entity embedding.}
\label{kg_embedding}
\end{figure*}
\begin{itemize}
\item \textbf{low-quality filter}:
We filter low-quality entities via some basic rules, such as blacklist, authority, hotness, importance and so on.
\item \textbf{low-frequency filter}:
We  filter low-frequency entities. 
\item \textbf{high-frequency sub-sampling}:
We make sub-sampling to those high-frequency entities. 
\item \textbf{shuffle}: 
We shuffle all samples.
\end{itemize}

Apart from user clicked data, we construct millions of query-entity relevant pairs at the semantic level, which are very important for the model to learn the query's semantic representation. Finally, we generate billions of query-entity pairs and about one thousand billion unique queries.

\begin{table}[h]\centering
        %\fontsize{8.5}{10}\selectfont
    \begin{tabular}{c||c|c|c|c}
    \hline
 \textbf{Method}&\textbf{P@1}&\textbf{P@10}&\textbf{P@20}&\textbf{P@30}\\
      \hline
      \hline
    DNN&6.53&28.29&38.83&53.79\\ 
      +ngram&7.25&30.76&41.57&56.49\\ 
   att-BiLSTM&7.34&30.95&41.56&56.02\\ 
        
     \hline
    \end{tabular}
    \caption{The offline comparison results of different methods in large-scale, real-world search logs of a widely used commercial web search engine.}
    
    \label{table5}
\end{table}

\subsection{Evaluation Metric}

To evaluate the effectiveness of different methods, we use Precision@M following \cite{zhu2018learning}. Derive the recalled set of entities for a query $u$ as $P_u (|P_u| = M)$ and the query's ground truth set as $G_u$. Precision@M  are:
 \begin{equation}
 \operatorname{Precision} @ M(u)=\frac{\left|\mathcal{P}_{u} \cap \mathcal{G}_{u}\right|}{M}
\end{equation}

\subsection{Offline Evaluation}
To evaluate the performance of our  model, we compare its performance with various baseline models. From unseen and real online search click log, we collect millions of query-entity pairs as our test set (ground truth set). The evaluation results are shown in Table \ref{table5}:      \textbf{DNN}  \cite{covington2016deep} is the  base method with a DNN encoder;  \textbf{+ngram}  is method adding ngram features; \textbf{att-BiLSTM}  is  our method  with BiLSTM  encoder with attention mechanism. The DNN \cite{covington2016deep} is a very famous recommendation baseline and  we re-implement the algorithm and modify the model  for entity recommendation setting. Note that, there are  no other baselines of  entity recommendation for  complex queries with no entities at all. \textbf{att-BiLSTM} is slightly better than \textbf{+ngram}. The reasons are mainly that a certain percentage of queries is without order and ngram is enough to provide useful information.

Our approach achieves the comparable results in the offline evaluation. 
These results indicate that our method benefits a lot from joint representation learning in queries and entities.  Note that, we learn the embedding of queries and entities with random initialization.  We believe the performance can be further improved by adopting more complex sentence encoder such as BERT\cite{devlin2018bert} and XLNet\cite{yang2019xlnet} and inductive bias from structure knowledge\cite{wang2018dkn} to  enhance the entity  representation, which we plan to address in future work.

\subsection{Online A/B Test}

We perform large-scale online A/B test to show how our approach on entity recommendation helps with improving the performance of recommendation in real-world applications.   We first retrieve candidate entities by matching queries, then we rank candidate entities by a click-through rate (CTR) prediction model and Thompson sampling. The ranked entities are pushed to users in the search results of Alibaba  UC Browser. For online A/B test, we split users into buckets. We observe and record the activities of each bucket for seven days. 
\begin{figure*}
\centering
\includegraphics [width=1\textwidth]{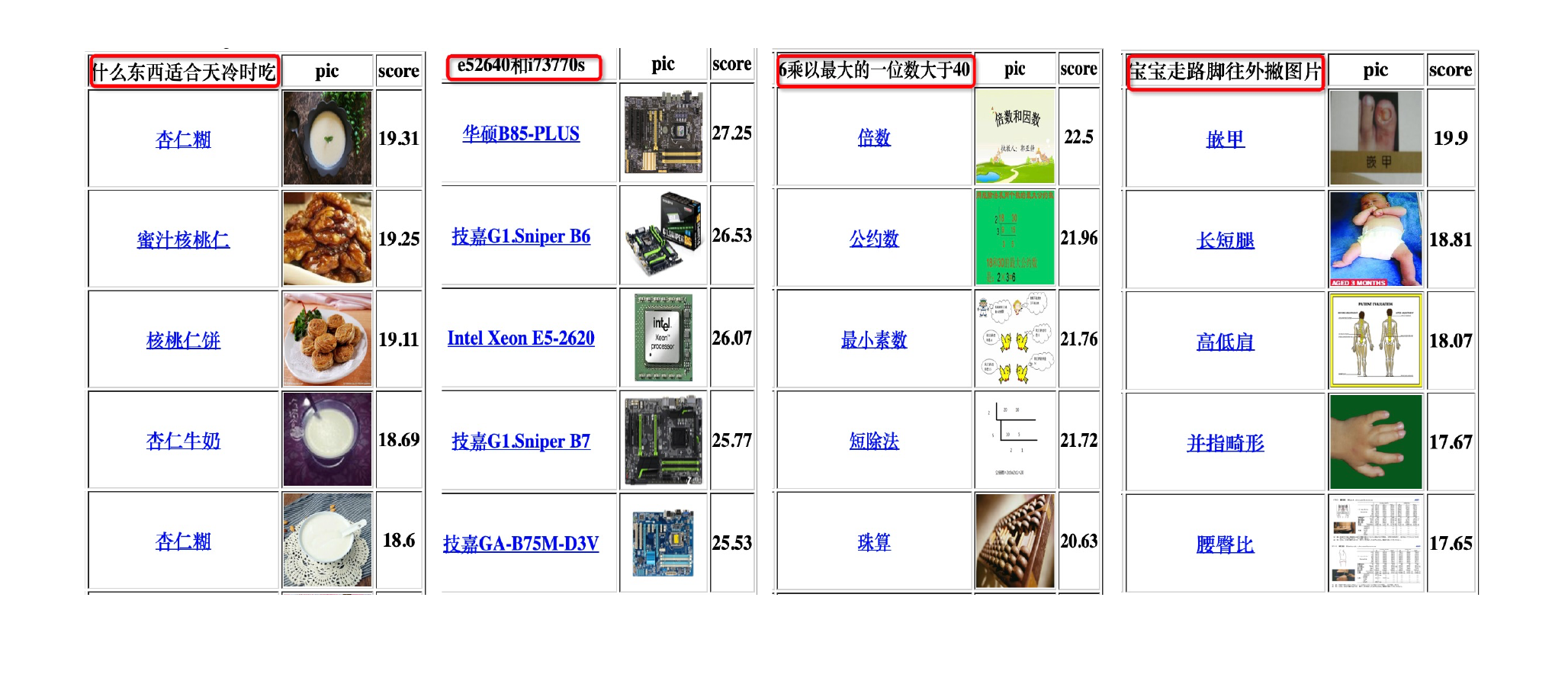}
\caption{Entity recommendation results from  complex and diverse queries.}
\label{case}
\end{figure*}
\begin{figure}
\centering
\includegraphics [width=0.4\textwidth]{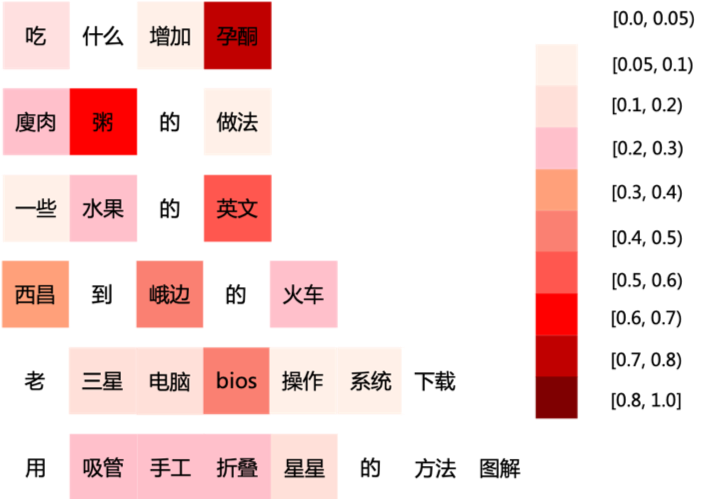}
\caption{Attention weights visualization of  six random queries from  search log.}
\label{attention}
\end{figure}

We select two buckets with highly similar activities. For one bucket, we perform recommendation without the deep collaborative match model. For another one, the deep collaborative match model is utilized for the recommendation. We run our A/B test for seven days and compare the result. The page view  (PV) and  click-through rate (CTR) are the two most critical metrics in real-world application because they show how many contents users read and how much time they spend on an application. In the online
experiment, we observe a statistically significant  CTR gain (\textbf{5.1\%}) and PV (\textbf{5.5\%}).   These observations prove that the deep collaborative match for entity recommendation greatly benefits the understanding of queries and helps to match users with their potential interested entities better. With the help of a deep collaborative match, we can better capture the contained implicit user's need in a query even if it does not explicitly have an entity.  Given more matched entities, users spend more times and reading more articles in our search engine.

\subsection{Qualitative Analysis}

We  make a qualitative analysis of the entity embeddings learned from scratch. Interestingly, we find that our approach is able to capture the restiveness of similar entities.  As Figure \ref{kg_embedding} shows, the entities "Beijing University," "Fudan University"   are similar to the entity "Tsinghua University." Those results demonstrate that our approach's impressive power of representation learning of entities\footnote{We do not have ground truth of similar entities so we cannot make quantitative analysis}.  It  also indicates that  the  text  is really  helpful in  representation learning in knowledge graph.

We also  make a qualitative analysis of the query embeddings.  We find that our approach generates more discriminate query embedding for entity recommendation due to the attention mechanisms.   Specifically, we randomly selected six queries from the search log and then visualize the attention weights, as shown in Figure \ref{attention}.  Our approach is capable of emphasizing those relative words and de-emphasizing those noisy terms in queries which boost the performance.

\subsection{Case Studies}
We give some examples of how our deep collaborative matching takes effect in entity recommendation for those complex queries.  In Figure \ref{case}, we display the most relative entities that are retrieved from the given queries.    We observe  that (1) given the  interrogative query "what food is good for cold weather", our model is able to understand  the meaning of query and get the most relative entities "Grain nutrition powder", "Almond milk"; (2) our model is able to handle short queries such as "e52640 and i73770s" which  usually do not have  the syntax of a written language or contain little signals for statistical inference; (3) our model is able to infer some queries such as "multiply six by the largest single digit  greater than fourth" that need commonsense  "number" is "mathematical terms"  which demonstrate the generalization  of our approach; (4) our approach can also handle  multi-modal queries "the picture of baby walking feet outside"  and get promising results although  in recent version of our model we do not consider the image representation in entity recommendation,   which indicates that our approach can  model the   presentation of  queries which reveal  the implicit need of users.  We believe the multi-modal  information (images) will further boost the performance which will be left for our future work.

\subsection{Conceptualized Entity Recommendation}

\begin{figure}
\centering
\includegraphics [width=0.35\textwidth]{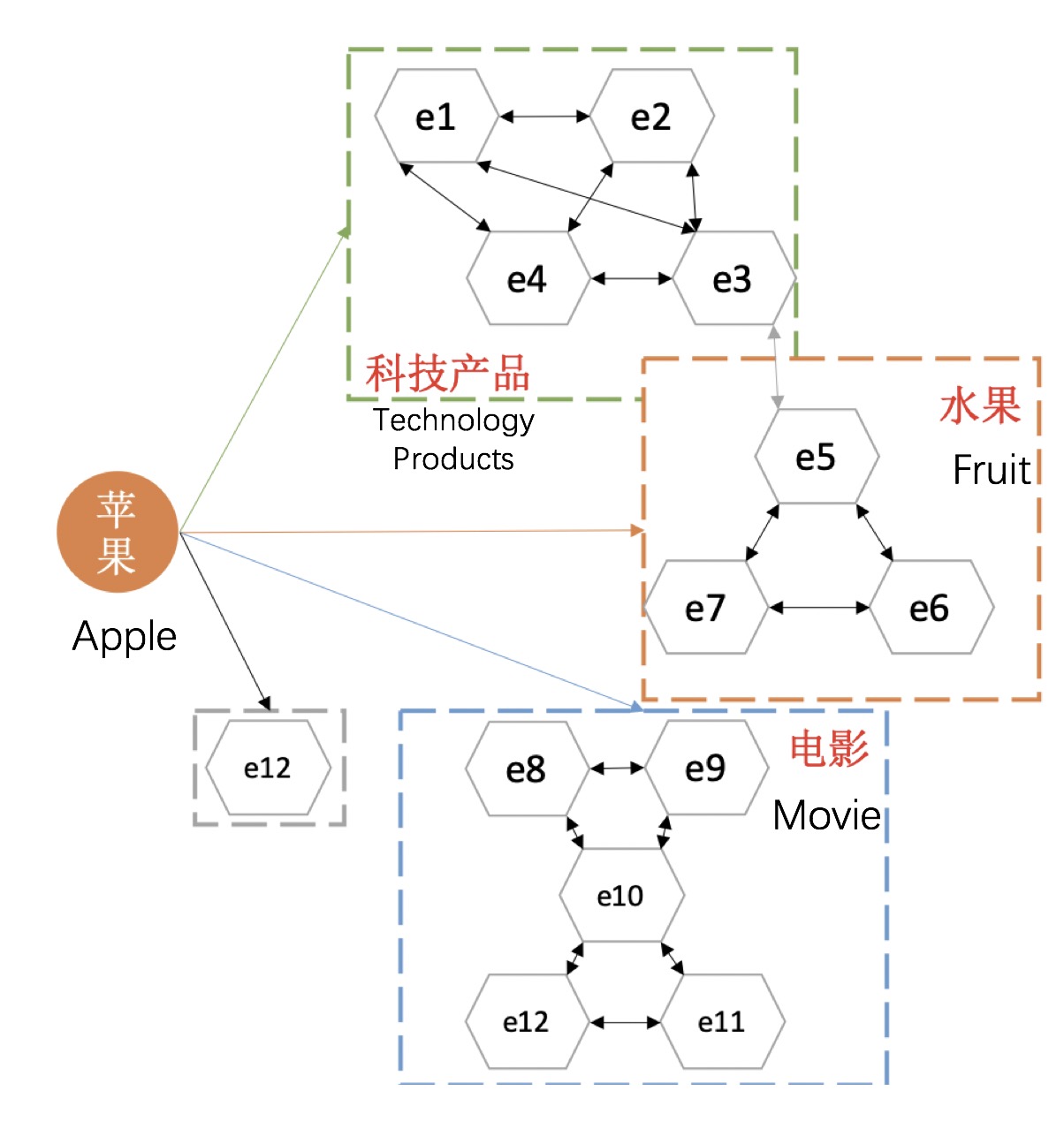}
\caption{Multiple concepts of an entity.}
\label{product1}
\end{figure}
\begin{figure}
\centering
\includegraphics [width=0.4\textwidth]{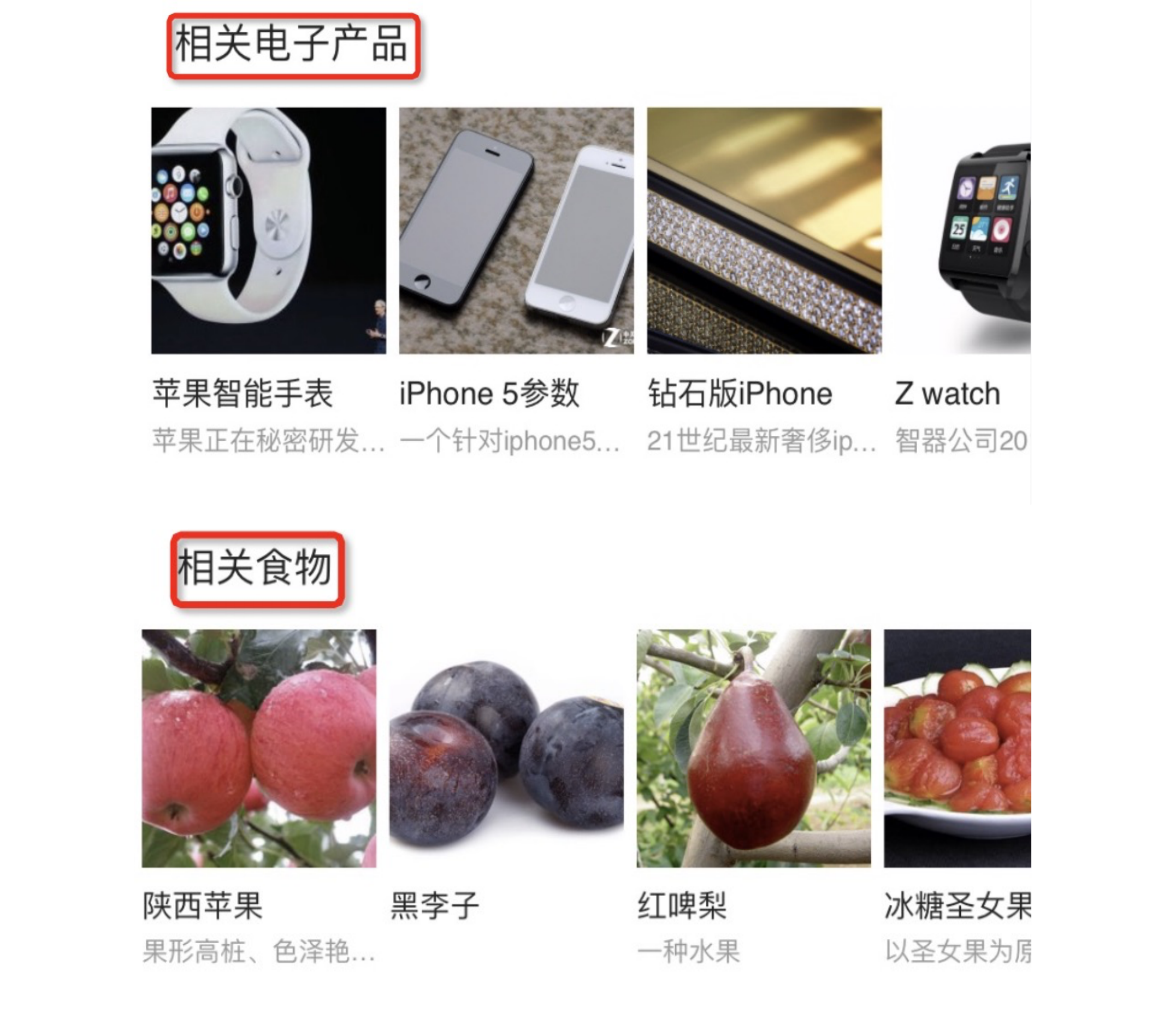}
\caption{Conceptualized multi-dimension entity recommendation. }
\label{product2}
\end{figure}
In the entity recommendation system,  each entity may have different views.  For example, when recommending entities relative to "apple", it may represent both "fruits" and "technology products" as the Figure \ref{product1} shows.   Actually, different users have different intentions. To give a better user experience, we  develop the conceptualized multi-dimensional  recommendation shown in Figure \ref{product2}. To be specific,  we utilize the concepts of candidate entities to cluster the  entities in the same group  to give a better visual display.  Those concepts are retrieved from our cognitive concept graph.  Online evaluation shows that conceptualized multi-dimensional  recommendation has the total coverage of 49.8\% in entity recommendation and also achieve more than 4.1\%  gain of CTR.

\section{Conclusion}

In this paper, we study the problem of context modeling for
improving entity recommendation. To this end, we develop a  deep collaborative match model that learns representations from complex and diverse queries and entities. We evaluate our approach using large-scale, real-world search logs of a widely used commercial search engine. The experiments demonstrate that our approach can significantly improve the performance of entity recommendation.

Generally speaking, the knowledge graph and cognitive concept graph can provide more prior knowledge in query understanding and entity recommendation.  In the future, we plan to explore the following directions: (1) we may combine our method with structure knowledge from knowledge graph and  cognitive  concept graph;  (2) we may combine rule mining and knowledge graph reasoning technologies to enhance the interpretability of entity recommendation; (3) it will be promising to apply  our method to other industry applications  and further adapt to other NLP scenarios.%. 

\section*{Acknowledgments}
We would like to thank colleagues of our team - Xiangzhi Wang, Yulin Wang, Liang Dong, Kangping Yin, Zhenxin Ma, Yongjin Wang, Qiteng Yang, Wei Shen, Liansheng Sun, Kui Xiong, Weixing Zhang and Feng Gao for useful discussions and supports on this work. We are grateful to our cooperative team - search engineering team. We also thank the anonymous reviewers for their valuable comments and suggestions that help improve the quality of this manuscript.

\
\bibliographystyle{ACM-Reference-Format}
\bibliography{sample-base}

\end{document}